# Status of the ILC Main Linac BPM R&D

M. Wendt[1], T. Lefevre[2], C. Simon[3], S. Vilalte[4] *

1 – Fermilab – AD-Instrumentation
Batavia, IL – U.S.A.

2 – CERN – CTF3
Geneva – Switzerland

3 – CEA Saclay – DSM/Irfu/SACM
Gif-sur-Yvette – France

4 – LAPP – CTF3 Group
Annecy – France

An introduction and the status of R&D activities for a high-resolution, "cold" beam position monitor (BPM) and the related read-out electronics are discussed [1]. Two different BPM detector concepts, to be attached to the SC quadrupole and located inside the ILC cryomodule, are currently under investigation: A resonant dipole-mode cavity-style BPM pickup, developed at Fermilab, and a re-entrant resonant coaxial waveguide BPM, designed by CEA-Saclay. While the 1.5 GHz dipole-mode cavity BPM is still in the R&D phase, the re-entrant BPM has already passed first beam tests, including its read-out system. Furthermore, the LAPP group is developing radiation tolerant digital read-out systems, which are tested at the CLIC test facility (CTF).

## 1 Introduction

Generation and transport of low emittance beams are crucial ILC operation aspects, required to provide a high luminosity at the IP. In order to preserve the low vertical beam emittance (0.04 mm mrad) along the ~11 km path through the SCRF Main Linac (ML), Beam Position Monitors (BPM) are installed in every 3$^{rd}$ cryomodule. These BPMs will be used to monitor the beam orbit, thus allow steering the beam on a dispersion-free, "golden" orbit. The ML BPM pickup principle is based on RF signal detection of the EM-field of the bunched beam, and is rigidly flanged to the superconducting (SC) quadrupole unit. These 280 "cold" BPMs are the only beam instruments inside the cryogenic sections in each of the Main Linac, therefore a high reliability of the ML BPM systems is mandatory.

Some of the requirements of the cold BPMs are more "exotic" than usual, because of the cryogenic operation temperature, and the neighborhood to the 1.3 GHz SCRF accelerating structures:

- The real estate for installation is limited to ~200 mm length and ~250 mm diametric size. The beam pipe aperture is circular, having 78 mm diameter.
- The BPM has to operate under ultra-high vacuum (UHV) conditions, and in a cryogenic environment at a temperature of ~2...10 K.
- A cleanroom class 100 certification is required to prevent pollution of the nearby SC cavities.
- A single bunch (bunch-by-bunch, i.e. < 350 ns measurement time) resolution of < 1

* This work was supported by Fermi National Accelerator Laboratory, operated by Fermi Research Alliance, LLC under contract No. DE-AC02-07CH11359 with the United States Department of Energy.

µm is required to preserve the low vertical emittance by applying dispersion-free orbit correction methods. This high resolution also allows for troubleshooting and diagnostics, e.g. spot sources of beam jitter, but can be guaranteed only nearby (± 1...2 mm) the electrical center of the BPM pickup.
- The absolute alignment error between electrical center of the BPM and magnetic center of the corresponding quadrupole should be < 300 µm. Larger errors will force to operate the BPM at an unwanted high "offset", and therefore reduce the resolution.
- A position linearity (scale error) of 2 % is requested, and was discussed, but not confirmed. A 5 % scale error seems to be more realistic (opinion of the author).
- The read-out system has to provide a sufficient dynamic range to operate over the full scale of beam displacements (0...±35 mm), as well as a reasonable range of bunch intensities (0.1...3.2 nC).

There are currently two BPM proposals to address the requirements, both based on a resonant beam pickup:
- A common-mode (CM) free dipole-mode cavity BPM pickup
- A re-entrant coaxial waveguide BPM

## 2  CM-"free" Cavity BPM

### 2.1  Principle of Operation

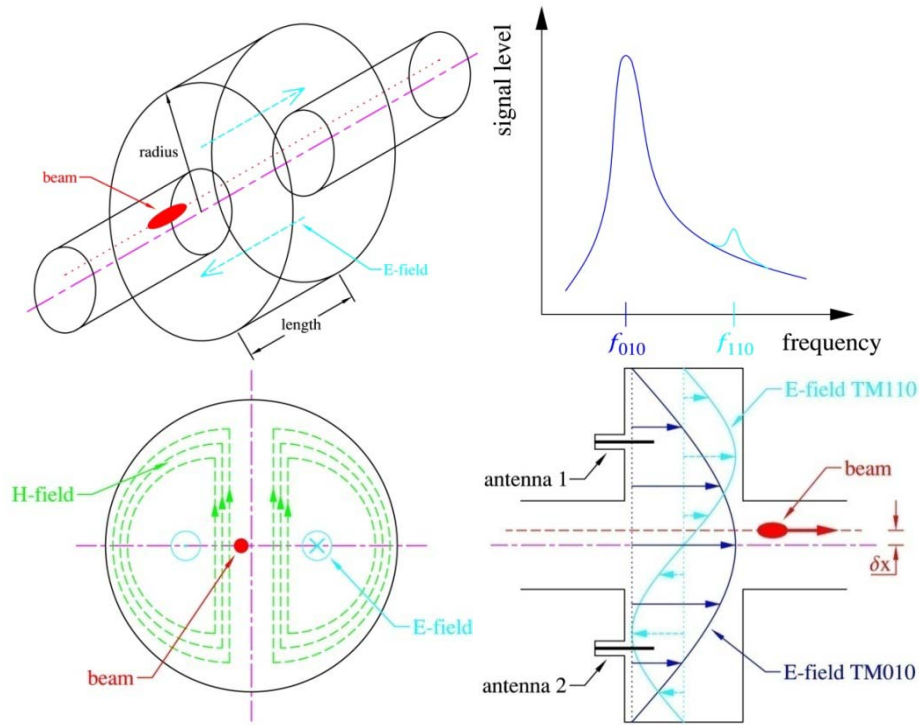

**Figure 1:** Principle of a "pillbox" cavity BPM

A cylindrical "pillbox" having conductive (metal) wall dimensions of radius $R$ and length $l$ resonates at eigenfrequencies:

$$f_{mnp} = \frac{1}{2\pi\sqrt{\mu_0 \varepsilon_0}} \sqrt{\left(\frac{j_{mn}}{R}\right)^2 + \left(\frac{p\pi}{l}\right)^2}$$

This resonator can be utilized as passive, beam driven cavity BPM by assembling it into the vacuum beam pipe. A subset of these eigenmodes is excited by the beam, for use as BPM the lowest transverse-magnetic dipole mode $TM_{110}$ is of interest. Its

$$E_z = C J_1\left(\frac{j_{11} r}{R}\right) \cos\phi \, e^{j\omega t}$$

field component couples to the beam, with almost linear dependence to the beams displacement $r$, and beam intensity (hidden in the constant $C$).

As Fig. 1 illustrates, the $TM_{110}$ dipole mode vanishes as the beam moves exactly through the center of the cavity pickup (r=0). A set of capacitive coupling pin antennas can be used to sense this dipole mode displacement signal of frequency $f_{110}$. Compared to broadband BPM pickups, e.g. button- or stripline style, the strength of the beam excited eigenmode depends on the cavity shape. For a simple cylindrical cavity BPM the corresponding $R_{sh}/Q$ (shunt impedance over Q-value) typically has much higher values, compared to the transfer impedance of a broadband BPM, i.e. has higher beam displacement sensitivity and therefore a much greater potential to function as a high resolution BPM.

Unfortunately this simple, basic pillbox cavity BPM setup has several issues, e.g.:
- $TM_{010}$ common mode suppression
- Cross talk between horizontal and vertical polarization of the dipole mode
- Transient response: loaded-Q and single bunch response
- Wakepotential and heat-load

which require some conceptual modifications to exploit the full resolution potential.

## 2.2 A cold CM-"free" cavity BPM for the ILC

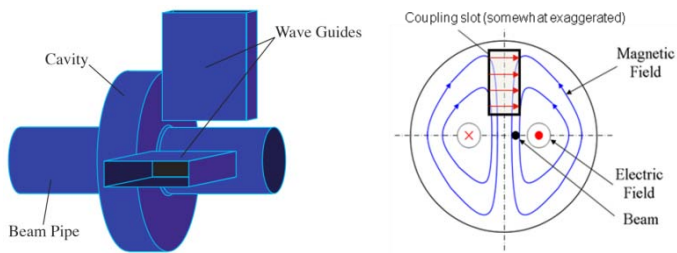

**Figure 2:** Waveguide-loaded CM-free cavity BPM (courtesy of Sean Walston).

The suppression of the $TM_{010}$ monopole mode is a key element to maximize the resolution of a cavity BPM. A set of two or four symmetrically arranged waveguides, coupled through slots to the beam excited cavity resonances, act as very efficient $TE_{01}$-mode high-pass filter (see Fig. 2.):

$$f_{010} \;<\; f_{10} = \frac{1}{2a\sqrt{\varepsilon\mu}} \;<\; f_{110}$$

Different cavity-waveguide coupling solutions and arrangements have been successfully tested in warm environments at SLAC, KEK and elsewhere, achieving sub-micrometer resolutions down to <10 nm. A cold L-Band cavity-BPM development at Fermilab tries to address all the discussed issues, details see Fig. 3. The waveguides are separated by ceramic-filled slot windows to simplify the cleaning procedure. Both, the $f_{110}$ = 1.468 GHz dipole mode, and the $f_{010}$ = 1.125 GHz monopole mode are read-out, thus there is no need of a reference signal. A prototype is currently in the assembly process, and will be tested soon (RF test stand and beam). A read-out system, based on an image rejection analog down-converter and a VME digital receiver for base-band detection is under development, with the focus on linearity / high dynamic range.

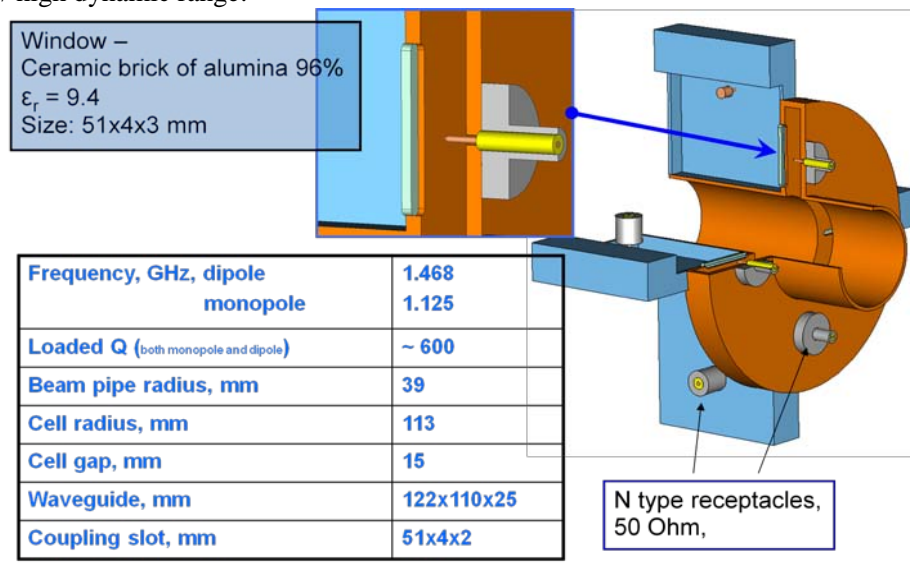

**Figure 3:** Cold L-Band cavity BPM development

## 3 Re-entrant BPM

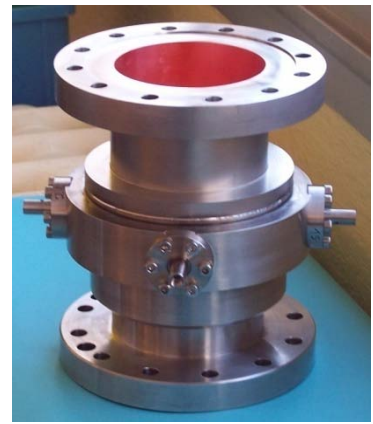

A re-entrant cavity BPM in Fig. 4 was specially designed to be connected to superconducting cavities which are particularly sensitive to dust particle contamination, and care must be taken to avoid introducing any source of such contamination.

The main features of this monitor are the small size of the RF cavity (170 mm), a large aperture (78 mm), an excellent linearity, a time resolution around 40 ns, and the beam charge measurement. A first prototype has already proved its operation at cryogenic temperature inside a cryomodule. Several cryogenic and vacuum tests (thermal shock) were carried out on the RF feedthroughs with success. As the BPM is designed to be used in a clean environment, twelve holes of 5 mm diameter are drilled at the end of the re-entrant part for more effective cleaning.

**Figure 4:** Re-entrant cavity BPM

The signal processing uses a single stage downconversion to obtain Δ/Σ and is shown in Fig. 5. It is composed of standard RF components: hybrid couplers, phase shifters, filters, isolators, and mixers. A printed circuit board (PCB) with low cost surface mount components will be designed to reduce cost.

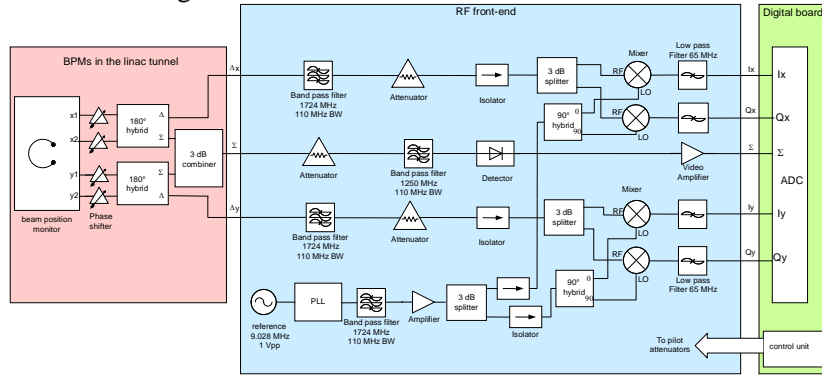

**Figure 5:** Re-entrant BPM read-out electronics

A prototype, installed in a warm section of the FLASH accelerator has been qualified with beam, achieving 4 µm resolution (Fig. 6) with 1 nC, limited only by the electromagnetic contamination in the experimental hall, over a dynamic range of ±5 mm and the possibility to do some bunch to bunch measurements. Those results are quite similar to the theoretical resolution calculated around 3.65 µm with 10 mm beam offset. If the dynamic range is limited to ±1 mm and an amplifier with a gain around 18 dB and a noise figure around 3.8 dB is added in the RF signal processing channel, this system gets a simulated resolution better than 1 µm.

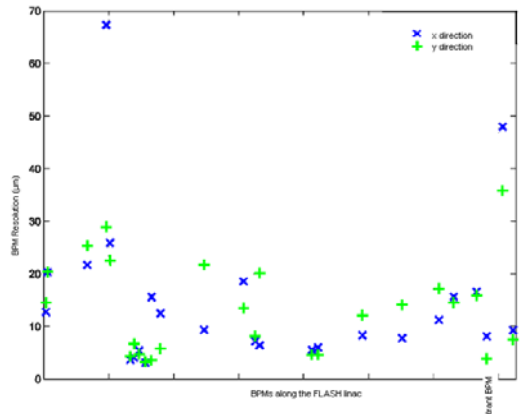

**Figure 6:** Resolution calculated for the re-entrant BPM installed in the warm part of the FLASH linac

Thirty cold re-entrant cavity BPMs will be installed in XFEL cryomodules and the next prototype will be installed in March 2009 in an XFEL cryomodule.

A second re-entrant BPM with an aperture of 18 mm and a large frequency separation between monopole and dipole modes, as well as a low loop exposure to the electric fields is developed for the CLIC Test Facility (CTF3) probe beam CALIFES at CERN. It is operated in single bunch and multi-bunches modes. The signal processing electronics is, too, based on downconversion. The resolution, in the single bunch mode, was calculated around 3 μm with a dynamic range of ± 5 mm and around 0.1 μm with a dynamic range of ± 0.1 mm.

## 4  Radiation Tolerant Read-out Electronics

In the CLIC test facility, two types of inductive pick-ups are used: BPMs from CERN (vertical and horizontal electrodes) and BPIs from INFN Frascati (45° tilted electrodes).
Until the transfer line 1, pick-ups signals were digitized far away from the accelerator using long analog cables and commercial VME boards. BPM signals were already preamplified and analog processed (horizontal, vertical and current) in the accelerator while BPI raw signals were digitized.
In order to reduce the cost of expensive analog cables between pick-ups and digitizers and because an acquisition close to the beam is a future beam diagnostic key, LAPP developed a radiation tolerant readout chain including:
   A preamplifier/processing analog module, compatible with the two types of pick-ups.
   An acquisition board (DFE) based on a 512Msps analog memory and a down-sampling 12 bits ADC.
The data collection corresponds to a 500ns window observed in the control room.
DFE boards are grouped by 4 in crates along the accelerator. The data transmission is daisy-chained and a specific board distributes the timings to the crate and calibration current pulses to the pick-up. So it allows a reduction by four of the number of cables and by 3 of the final cost of the electronics.
All combinations of pick-ups – preamplifiers – DFE are present in the machine.
In its final layout configuration, the CTF3 will include 47 analog modules and 46 DFE boards, all distributed in 12 crates. Two computers collect the data in a gallery above the accelerator. At the moment, 31 are used for the commissioning of the machine.
In 2009, an upgraded optical linked version with a simplified architecture will be developed. It should allow to eliminate the cables and will be dedicated to a larger accelerator as CLIC: rare accesses from the surface, high number of channels, low cost, low consumption, autonomous power supplies and calibrations.